\documentclass[conference]{IEEEtran}
\IEEEoverridecommandlockouts
\usepackage{cite}
\usepackage{amsmath,amssymb,amsfonts}
\usepackage{algorithmic}
\usepackage{graphicx}
\usepackage{textcomp}
\usepackage{xcolor}
\usepackage[most]{tcolorbox}
\usepackage{marginnote}
\definecolor{C1}{HTML}{1F77B4}

\DeclareGraphicsExtensions{.pdf,.png,.jpg,.mps,.eps,.ps}
\usepackage{amsmath}
\usepackage{amsfonts}
\usepackage{mathtools}

\usepackage{forloop}
\newcommand{\defvec}[1]{\expandafter\newcommand\csname v#1\endcsname{{\mathbf{#1}}}}
\newcounter{ct}
\forLoop{1}{26}{ct}{
    \edef\letter{\alph{ct}}
    \expandafter\defvec\letter
}

\forLoop{1}{26}{ct}{
    \edef\letter{\Alph{ct}}
    \expandafter\defvec\letter
}

\usepackage{bm} 

\newcommand{\vlambda}{\bm{\lambda}}
\newcommand{\vtheta}{\bm{\theta}}

\DeclareMathOperator*{\E}{\mathbb{E}} 
\DeclareMathOperator*{\var}{var}
\DeclareMathOperator*{\cov}{cov}
\DeclareMathOperator*{\tr}{Tr}
\DeclareMathOperator{\diag}{diag}

\DeclarePairedDelimiter{\norm}{\lVert}{\rVert}

\newcommand{\inv}{^{-1}}
\newcommand{\field}[1]{\ensuremath{\mathbb{#1}}}

\newcommand{\reals}{\field{R}}

\newcommand{\trp}{^\top} 

\newcommand{\iidsample}{\stackrel{iid}{\sim}}
\newcommand{\normalDist}{\mathcal{N}}
\newenvironment{noticebox}[1][b]{
    \enlargethispage{2\baselineskip}
    \begin{figure}[#1]
    \footnotesize
}{
    \end{figure}
}

\def\BibTeX{{\rm B\kern-.05em{\sc i\kern-.025em b}\kern-.08em
    T\kern-.1667em\lower.7ex\hbox{E}\kern-.125emX}}

\begin{document}
\title{Quantifying Signal-to-Noise Ratio in Neural Latent Trajectories via Fisher Information\\
\thanks{Supported in part by NIH RF1 DA056404, Portuguese Recovery and Resilience Plan (PPR) 62, and Funda\c{c}\~{a}o para a Ci\^{e}ncia e a Tecnologia (FCT) project UIDB/04443/2020.}
}

\author{\IEEEauthorblockN{Hyungju Jeon}
\IEEEauthorblockA{\textit{Champalimaud Research} \\
\textit{Champalimaud Foundation}\\
Lisbon, Portugal \\
hyungju.jeon@research.fchampalimaud.org}
\and
\IEEEauthorblockN{Il Memming Park}
\IEEEauthorblockA{\textit{Champalimaud Research} \\
\textit{Champalimaud Foundation}\\
Lisbon, Portugal \\
memming.park@research.fchampalimaud.org }
}

\maketitle

\begin{noticebox}
    This article is accepted for publication in the 2024 European Signal Processing Conference (EUSIPCO) 
\end{noticebox}

\begin{abstract}
    Spike train signals recorded from a large population of neurons often exhibit low-dimensional spatio-temporal structure and modeled as conditional Poisson observations.
    The low-dimensional signals that capture internal brain states are useful for building brain machine interfaces and understanding the neural computation underlying meaningful behavior.
    We derive a practical upper bound to the signal-to-noise ratio (SNR) of inferred neural latent trajectories using Fisher information.
    We show that the SNR bound is proportional to the overdispersion factor and the Fisher information per neuron. Further numerical experiments show that inference methods that exploit the temporal regularities can achieve higher SNRs that are proportional to the bound.
    Our results provide insights for fitting models to data, simulating neural responses, and design of experiments.

\end{abstract}

\begin{IEEEkeywords}
    Fisher information, neural code, latent neural dynamics, Poisson likelihood
\end{IEEEkeywords}

\section{Introduction}
Inferring low-dimensional descriptor (neural states) and their temporal evolution (neural trajectories) from population recordings of neurons arise in many applications such as brain-machine interfaces, study of neural code and computation, and understanding the neural basis of meaningful behavior~\cite{Degenhart2020-gu,Zhao2019b}.
In the field of neuroscience, inferring neural states and trajectories often relies on dimensionality reduction techniques with various assumptions~\cite{Yu2009-qp,Zhao2016a,Zhao2019a,Dowling2024b}. While the recent advances in neural recording technology and computational methods have enabled large scale recordings, allowing more accurate inferences, the success of these method hinges not just on the quantity of data but, more importantly, on the overall informativeness of the recorded population regarding the latent neural trajectories.
One key aspect is the signal-to-noise ratio (SNR), which quantifies the amount of useful information relative to the noise in the data. Determining the SNR of the inferred latent trajectories can guide the design of experiments by providing insights into various experimental parameters such as the number of neurons to record and the temporal resolution we would need to observe to achieve the target precision~\cite{Gao2015-yd}. SNR analysis also provides a principled way to evaluate statistical neuroscience algorithms.

Previous studies have utilized Fisher information to establish bounds on mutual information in neural populations \cite{Brunel1998-pr}. \cite{Bartolo2020-zd} analyzes the correlated noise that aligns with the Fisher information matrix. \cite{Bethge2002-ji} shows SNR regimes  where the Fisher information fails to accurately predict decoding performance. \cite{Beck2011-kt} employs linear Fisher information to quantify the correlation in log-linear Poisson neurons. \cite{Wei2016-mi} examines the optimal tuning curves of neurons using Fisher information. 
However, previous works were in the context of stimulus encoding where the so-called ``noise correlation'' played an important role\cite{Moreno-Bote2014-ie}. When considering the latent states as the signal itself\cite{Hurwitz2021-pf}, there is no distinction between correlation due to the trial-to-trial variability and the signal itself.

In this paper, we derive a practical upper bound to the SNR using Fisher information. Our model assumes that the firing rate of each neuron is a function of the low-dimensional latent state, and that the observed neural responses follow a Poisson distribution.
We apply our model to a non-human primate neural population recording data~\cite{Pei2021b}. We find that the marginal overdispersion measured by Fano factor~\cite{Charles2018} is a good proxy for the Fisher information in the log-linear Poisson model.
Our analysis extends prior works and can be applied without the full knowledge about the properties of the signal distribution. Another important distinction is that our analysis does not assume the stimulus to be constant for a long period of time, but rather dynamically varying.

\section{Derivation}\label{sec:generative}
\subsection{Log-linear observation model}
The population tuning curve $\vlambda: \reals^L \to \reals^N$ defines the firing rate of $N$ neurons as a function of the latent state $\vx \in \reals^L$. In the case of log-linear model the firing rate is given by,
\begin{align}\label{eq:loglinear}
    \vlambda(\vx) &= \exp\left(\vC\vx + \vb\right)
\end{align}
where $\vC \in \reals^{N \times L}$ is the loading matrix, $\vb \in \reals^N$ is the bias vector and $\exp(\cdot)$ is applied element-wise. Note the time index are omitted for brevity.
We assume the ``low-dimensional'' dynamics regime, where $L \ll N$\cite{Gao2015-yd}.
Derivative of the tuning curve for the $k$-th neuron with respect to the parameter of interest, the $i$-th latent state $\vx_i$, is,
\begin{align}\label{eq:loglinear:derivative}
    \frac{\partial \lambda_k}{\partial \vx_i}
        &= \lambda_k \frac{\partial}{\partial \vx_i} \left(\vC_k\vx + b_k\right)
        = \lambda_k C_{k,i}
\end{align}
The observed neural response $y_k$ is modeled as a conditionally Poisson distribution with rate $\lambda_k$, a discrete-time approximation of an inhomogeneous Poisson process:
\begin{align}\label{eq:poisson:log}
    \log p(y_k | \lambda_k) &= y_k \log \lambda_k - \lambda_k - \log(y_k!)
\end{align}
Partial derivative of observed neural response $y_k$ with respect to some arbitrary parameters $\vtheta$ is,
\begin{align}\label{eq:poisson:d}
    \frac{\partial \log p(y_k)}{\partial \vtheta} 
        &= \left(\frac{y_k}{\lambda_k} - 1\right)
            \frac{\partial \lambda_k}{\partial \vtheta}
\end{align}

Recall the definition of the Fisher information matrix for a random variable $y$ with respect to a parameter $\vtheta$:
\begin{align}\label{eq:fisher:def}
    I_{i,j}(\vtheta) &= \mathbb{E}_{y \mid \vtheta}
        \left[
            \frac{\partial}{\partial \theta_i} \log p(y)
            \frac{\partial}{\partial \theta_j} \log p(y)
        \right]
\end{align}
Plugging in the Poisson log-likelihood derivative \eqref{eq:poisson:d}, and setting the parameters $\vtheta$ to be our latent state $\vx$, the Fisher information matrix for the $k$-th neuron is, 
\begin{align}\label{eq:fisher:poisson}
    I_{i,j}^k(\vtheta)
    &= \E_{y_k \mid \vx}
        \left[
            \left(\frac{y_k}{\lambda_k} - 1\right)^2
            \frac{\partial \lambda_k}{\partial \vx_i}
            \frac{\partial \lambda_k}{\partial \vx_j}
        \right]\notag
    \\
    &=
    \frac{1}{\lambda_k(\vx)}
    \frac{\partial \lambda_k(\vx)}{\partial \vx_i}
    \frac{\partial \lambda_k(\vx)}{\partial \vx_j}
\end{align}
Plugging in the log-linear tuning curve \eqref{eq:loglinear:derivative}, we get the Fisher information matrix for the $k$-th neuron as a function of $\vx$:
\begin{align}\label{eq:fisher:loglinear:neuron}
    I_{i,j}^k(\vx)
    &=
        \lambda_k(\vx) C_{k,i} C_{k,j}
\end{align}
Since each neuron is conditionally independent given $\vx$, the Fisher information matrix for the entire population can be calculated as the sum of the Fisher information matrices for each neuron,
\begin{align}\label{eq:fisher:loglinear:pop}
    I_{i,j}^{\text{pop}}(\vx)
    &=
        \sum_{k=1}^N I_{i,j}^k(\vx)
    =
        \sum_{k=1}^N \lambda_k(\vx) C_{k,i} C_{k,j}
\\
    \bm{I}^{\text{pop}}(\vx)
        &=
            \vC\trp \diag(\vlambda(\vx)) \vC
\end{align}

\subsection{Upper bound on the Signal-to-Noise ratio}
The SNR is the ratio of the signal variance to the noise variance.
In this work, we aim to assess how accurately we can estimate the latent state $\vx$ from the spikes $\vy$?
Here, the signal of interest is the latent state $\vx$ and the noise arises from the Poisson spiking process.
Given an estimator $\hat{\vx}$, we can define the SNR as,
\begin{align}\label{eq:snr:def}
    \text{SNR} &= 
    \frac{\text{signal power}}{\text{noise power}}
    =
    \frac{\E[ \norm{\vx}^2]}{\E[\norm{\hat{\vx} - \vx}^2]}.
\end{align}
Note that the $R^2 = 1 - \text{SNR}\inv$ is the fraction of the variance of the latent state $\vx$ that is recoverable from the spikes $\vy$.

We can use Cramer-Rao bound of the covariance of any unbiased estimator $\hat{\vx}$~\cite{Bethge2002-ji}:
\begin{align}\label{eq:crbound:norm}
    \cov[\hat{\vx} \mid \vx] &\geq \bm{I}\inv(\vx)
   \\
   \tr\left(
           \cov[\hat{\vx} \mid \vx]
       \right)
   &\geq \tr 
       \left(
           \bm{I}\inv(\vx)
       \right)
\end{align}
which is a function of the true latent state $\vx$ and 
where the inequality denotes the positive semi-definite ordering.

Therefore, given a distribution over the latent states, we can take the expectation to 
get a lower bound on the mean squared error (MSE) of the estimator $\hat{\vx}$.
\begin{align}
    \E[\norm{\hat{\vx} - \vx}^2] &=
        \E[\E[\norm{\hat{\vx} - \vx}^2 \mid \vx]]
        \nonumber
        \\
        &\geq \E\left[\tr \left(\bm{I}\inv(\vx)\right) \right]
        \nonumber
        \\
        &= \E[\tr\left(\vC\trp \diag(\vlambda(\vx)) \vC\right)\inv]
        \label{eq:crbound:expectation}
\end{align}
We will use \eqref{eq:snr:def} to compute an upper bound of SNR of the latent state:
\
\begin{tcolorbox}[enhanced,colback=white,%
	colframe=C1!75!black, attach boxed title to top right={yshift=-\tcboxedtitleheight/2, xshift=-.75cm}, title=Fisher upper bound of SNR, coltitle=C1!75!black, boxed title style={size=small,colback=white,opacityback=1, opacityframe=0}, size=title, enlarge top initially by=-\tcboxedtitleheight/2, left=1pt]
\begin{align}
    \text{SNR} &\leq \frac{\E[ \norm{\vx}^2]}{ \E[\tr\left(\vC\trp \diag(\vlambda(\vx)) \vC\right)\inv]}
    \label{eq:snr:bound}
\end{align}
\end{tcolorbox}

\section{Results}
\subsection{1D latent state case}
In general, \eqref{eq:snr:bound} is not analytically tractable, however, we can still gain insight into the scaling of the SNR with simple $L = 1$ log-linear Poisson model.
We will assume that $x \iidsample \normalDist(0, 1)$.
\begin{align}
    \lambda_k(x) &= e^{c_k x + b_k}
    \\
    \E[\lambda_k(x)] &= e^{b_k} e^{\frac{1}{2} c_k^2}
    \\
    \var[\lambda_k(x)] &= e^{2b_k} (e^{2 c_k^2} - e^{c_k^2})
    \\
    \var[y_k] & = \E[\var(y_k \mid x)] + \var[\E[y_k \mid x]] \notag
    \\
    &= e^{b_k} e^{\frac{1}{2} c_k^2} + e^{2b_k} (e^{2 c_k^2} - e^{c_k^2})
\end{align}

The overdispersion can be measured with Fano factor~\cite{Charles2018}, which is the ratio of the variance to the mean firing rate.
Fano factor is $1$ for Poisson process and greater than $1$ for a mixture of Poisson processes.
The overdispersion is defined as the excess Fano factor relative to the unity.
\begin{align}
    \text{Fano factor} &= \frac{\var[y_k]}{\E[y_k]} \\
                       &= 1 + e^{b_k} (e^{\frac{3}{2} c_k^2} - e^{\frac{1}{2}c_k^2})\\
                       &\simeq  1 + e^{b_k} (c_k^2 + c_k^4) + O(c_k^6)
\end{align}
If we constrain the mean firing rate to be $\bar{\lambda}$, 
\begin{align}
    b_k = \log \bar{\lambda} - \frac{1}{2} c_k^2,
    &\qquad
    \lambda_k(x) = \bar{\lambda} e^{c_k(x - \frac{1}{2} c_k)}\\
    \text{Fano factor}    
    &=1 + \bar{\lambda} (e^{c_k^2} - 1)\label{eq:FF:1d}\\ 
    &\simeq 1 + \bar{\lambda} c_k^2 + O(c_k^4)
\end{align}
the overdispersion scales with $c_k^2$.

Note that the Fisher information is,
\begin{align}
    I^{\text{pop}}(x) &= \sum_k c_k^2 \lambda_k(x)
\end{align}
Thus, larger $c_k^2$ and $x$ that produces larger firing rate are more informative.
Under the mean firing rate constraint, we have,
\begin{align}
    I^\text{pop}(x) &= 
        \bar{\lambda}\sum_k
        c_k^2
        e^{c_k(x - \frac{1}{2} c_k)}
\end{align}
Given $x \iidsample \normalDist(0, 1)$, the expected Fisher information is a bound of SNR:
\begin{align}
    \E[I^\text{pop}(x)] &= 
        \bar{\lambda}\sum_k
        \left(
            c_k^2 
            e^{-\frac{1}{2} c_k^2}
        \right)
        \E[ e^{c_k x} ] \notag
    \\
        &=
        \bar{\lambda}
        \sum_k
        c_k^2
        \geq \text{SNR}
\end{align}
Note that it coincides with the overdispersion factor in \eqref{eq:FF:1d} up to the first order in $c_k$.
\subsection{Numerical experiments}
For more general cases, we numerically analyzed the behavior of SNR bound for various recording scenarios.
To simulate the observation model with a target SNR, the loading matrix $\vC$ and base firing rate $\vb$ are initialized using random values drawn from a normal distribution. Then the $\vC$ is scaled uniformly by a tuning strength $g$ to achieve the target SNR, and the base firing rate $\vb$ is adjusted to achieve the target mean firing rate of 10 Hz.
The latent state $\vx  \in \reals^{3 \times T}$ was generated from a Gaussian process.
\begin{figure}[h!bt]
    \includegraphics[width=\linewidth]{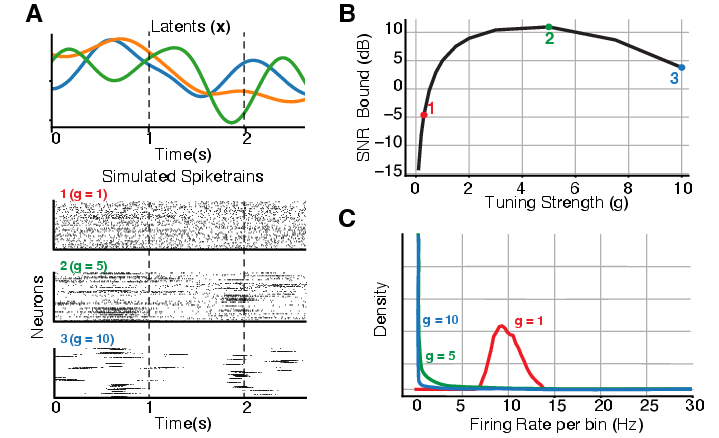}
    \caption{
    (A) Example latent trajectories ($L=3$) and simulated neural responses for varying tuning strength $g$.
    (B) SNR upper bound for different $g$.
    (C) Distribution of firing rate for different $g$.
    }
    \label{fig:numerical:tuning}
\end{figure}

Consistent with the results from the 1D case, in Fig.~\ref{fig:numerical:tuning} we observed that population of neurons with large loading matrix $\vC$ are more narrowly tuned to the latent $\vx$, more informative and thus provide higher SNR bound. However, the SNR bound is not monotonic with respect to the tuning strength due to mean firing rate constraint and homogeneous scaling in loading matrix. 
\subsection{Cortical data analysis}
In this section, we apply the theory to a neural population recording from a non-human primate.
We use a publicly available dataset from the Neural Latent Benchmark,
a neural population recording from dorsomedial frontal cortex (DMFC) while a monkey performs a time interval reproduction task~\cite{Pei2021b}.
We use the log-linear Poisson state space model with a Gaussian process assumption fit to the data~\cite{Zhao2016a}.
The parameters $\vC, \vb$ and the latents $\vx(t)$ were obtained using a recent advance in the variational inference algorithm that scales linearly in time~\cite{Dowling2023c}.
From the 54 simultaneously recorded neurons and $2.24 \times 10^6$ time bins, a 3-dimensional latent state space was recovered (for details see~\cite{Dowling2023c}).
We preprocess the inferred latent trajectories by centering and whitening the covariance matrix and simultaneously transforming the loading matrix such that the empirical mean of the $\ell_2$-norm of the latents is $L=3$.
\begin{figure}[!htb]
    \includegraphics[width=\linewidth]{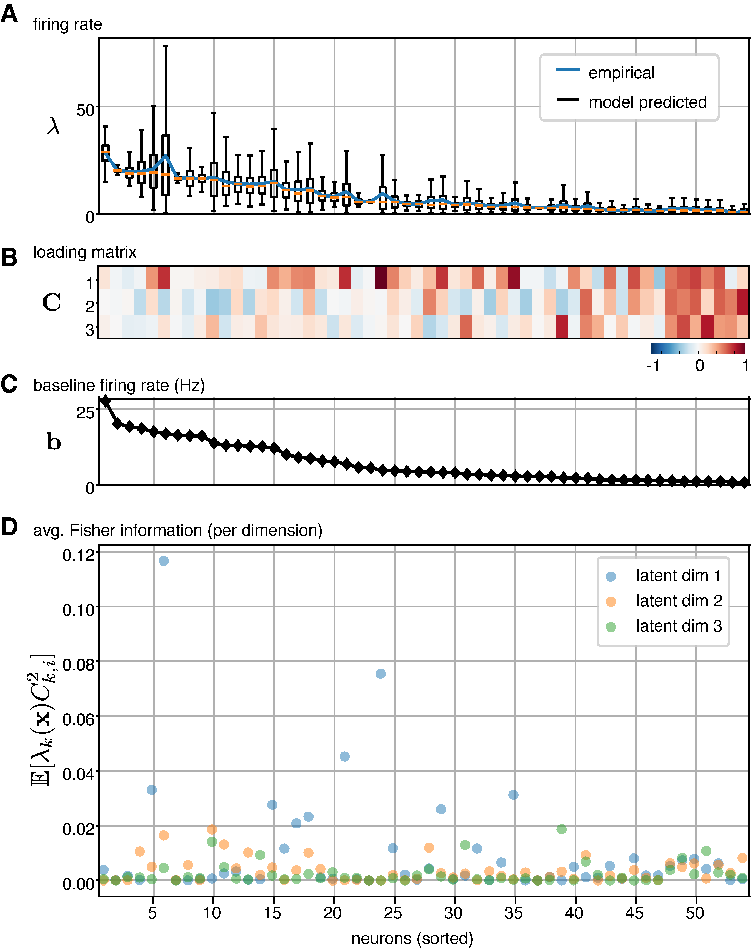}
    \caption{
        (A) Firing rate distribution of the neurons.
        Box plots are model predicted firing rate per 5 ms bin.
        Blue line is the average firing rate of the observed spikes in the training set across all time bins.
        Neurons are sorted in decreasing order of $\vb$.
        (B) The loading matrix $\vC$.
        (C) The baseline firing rate $\exp(\vb)/\Delta$.
        (D) Fisher information per neuron per latent dimension averaged over time.
    }
    \label{fig:dmfc:neuron}
\end{figure}
In Fig.~\ref{fig:dmfc:neuron}A, we observe that the recorded neurons have a broad firing rate distribution, and the model matches the heterogeneity well.
Most of the firing rate differences are explained by the difference in the bias (Fig.~\ref{fig:dmfc:neuron}C),
but variations in the firing rate for each neuron (with high Fano factor) corresponds to high magnitude of the loading matrix (Fig.~\ref{fig:dmfc:neuron}B).

The loading matrix $\vC$ shown in Fig.~\ref{fig:dmfc:neuron}B is not likely to be from a random normal matrix.
Rather, the loading matrix is sparse and has a few large magnitude entries.
Using the inferred parameters $\vC, \vb, \vx(t)$, we can compute the Fisher information matrix for each neuron at around the average rate of that neuron.
As shown in Fig.~\ref{fig:dmfc:neuron}D, the contribution to the Fisher information is dominated by a few neuron-latent pairs.
We can estimate the SNR of the latent state by computing the Fisher information matrix without assuming the normality of the latent state distribution by taking the empirical expectation of \eqref{eq:crbound:expectation} by averaging over time bins.
For the DMFC dataset, using the entire population, the SNR on average is $-7.60$ dB
(per dimension: $-6.76$, $-8.76$, $-7.00$ dB) per a $5$ ms bin.
This is a very low SNR, but consistent with the sparse firing of neurons in a $5$ ms bin.
Thankfully, the latent neural state is highly correlated over time, and this temporal smoothness can be exploited for the estimation, which is utilized by methods such as Gaussian process factor analysis (GPFA)~\cite{Yu2009-qp}, Poisson linear dynamical systems (PLDS)~\cite{Macke2011-ep}, and variational latent Gaussian processes (vLGP)~\cite{Zhao2016a}.

\begin{figure}[h!bt]
    \includegraphics[width=\linewidth]{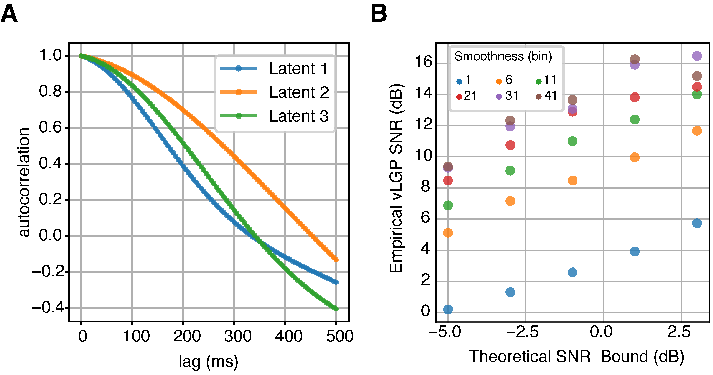}
    \caption{
	    (A) Autocorrelogram of the latent trajectories. Note that the time constant is longer than 100 ms. (B) Empirical SNR using vLGP inference on simulated latent trajectories with varying time scale.
    }
    \label{fig:vlgp_smoothing}
\end{figure}

As expected from the time scale of the behavior and cognitive task the subject was performing,
the time constant is long, indicated by the autocorrelation function of the estimated latent trajectories per dimension.
Therefore, the SNR can be improved if the inference can utilize the temporal smoothness.
Since the firing rate per bin is a linear function of time window, the Fisher information also scales linearly, improving SNR by $10 \log_{10}(T)$ dB with $T$ times longer time window.
Given the autocorrelation functions of the latents (Fig.~\ref{fig:vlgp_smoothing}A), we can expect the SNR to improve by approximately $10 \log_{10}(25) \approx 14$~dB by using window size of 25 ms.
To test this numerically, we generated the latent trajectories using Gaussian process with time scale parameter of 250 ms. We simulated neuronal firings using log-linear Poisson observations model with different tuning strength, hence the different SNR bound, and computed the empirical SNR of the vLGP inference result with varying time scale parameters. As shown in Fig.~\ref{fig:vlgp_smoothing}B, the SNR improves as the time scale increases, consistent with the theory.
\begin{figure}[htb]
    \includegraphics[width=0.9\linewidth]{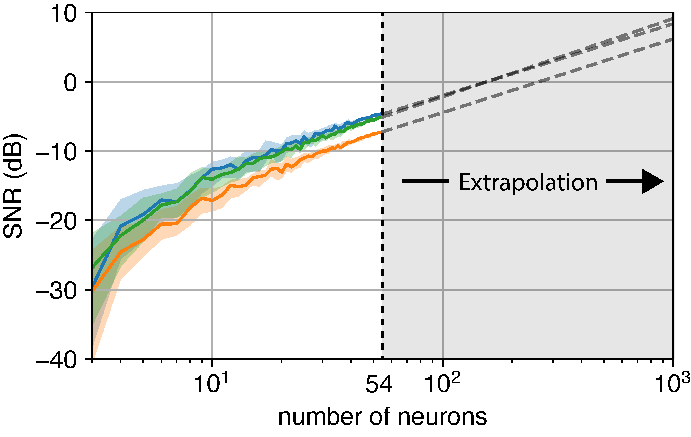}
    \caption{
        By resampling a subset of neurons, the SNR bounded by Fisher information is estimated
        as a function of the recorded population size up to 54 neurons (dotted lines).
	Each color corresponds to one dimension of the latent process in Fig.~\ref{fig:dmfc:neuron}.
        The dashed lines are power-law fit of the trend that extrapolates to 1000 neurons.
    }
    \label{fig:dmfc:extrapolation}
\end{figure}

Another way to improve the SNR is to record more neurons.
To answer the question of how many neurons we need to record simultaneously, we subsample the neurons and compute the SNR bound\cite{Stevenson2011-wk}.
In Fig.~\ref{fig:dmfc:extrapolation}, we show the SNR as a function of the number of neurons.
Under the assumption of independent and identical loading matrix entries for yet unobserved neurons,
we can extrapolate the SNR to 1000 neurons with a power-law fit.
General trend increases the SNR by around 12 dB at 20 times larger recording setup.

\begin{figure}[htb]
    \includegraphics[width=\linewidth]{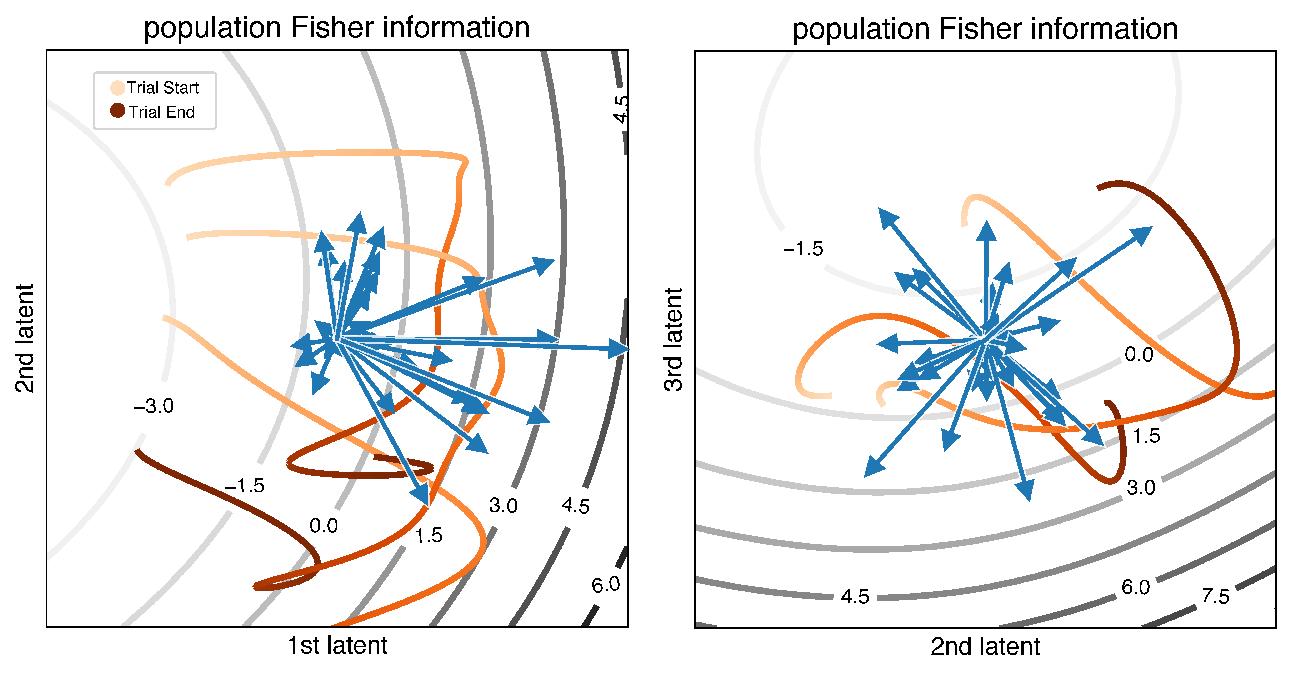}
    \caption{
        Population Fisher information visualized for 2-dimensional slices.
        For each pair of dimensions, $[-2, 2]^2$ area is visualized.
        The arrows are the rows of $\vC$ matrix (with a scale factor of 0.3 for visualization).
        The contour in the background is $\sum_{k} \lambda(\vx)_k C_{k,i}^2$.
        The colored trajectories are 3 randomly selected trials (each 300 time bins long).
    }
    \label{fig:dmfc:2x2}
\end{figure}

To get more insights into how each neuron is contributing to the Fisher information, and which part of the state space has better Fisher information,
we visualize 2D slices of the Fisher information as a function of the latent states (Fig.~\ref{fig:dmfc:2x2}).

Due to the uneven distribution of the rows of the loading matrix, there are directions in the latent space that are more readily estimated than others (Fig.~\ref{fig:dmfc:2x2} arrows and contour).
Since we pre-processed the latent state trajectories to have zero mean and unit variance, this anisometry may be a result of the inference algorithm.
Indeed, we do see that during each trial, particularly when the subject is involved in the timing task, the trajectories tends to reside in the region with higher Fisher information (Fig.~\ref{fig:dmfc:2x2} trajectory).

At the same time, due to the form of the log-linear Poisson Fisher information matrix, the Fisher information is low near the origin.
This has an interesting implication regarding neural state space geometry and neural codes.
For example, neural codes that restricted to a spherical manifold would be have benefits compared to a linear manifold that passes through the origin.
Interestingly, on a ring geometry, the tuning curves of individual neurons induced by the log-linear model becomes the familiar cosine tuning curves.

It is not only the neuroscientists and the neural engineers who are interested in recovering the latent states,
but other neurons and brain areas must also be able to access the latent states.
As the neural state space is bounded, and the neural codes tend to lie on curved manifolds.
In high-dimensional embedding space such as cortical population representations, these manifolds typically occupy regions away from the origin.
The usual Gaussian distribution prior on the latent states also has most of its probability mass away from the origin in higher dimensions, and dimensionality reduction techniques maximizes variance which also repels inferred states away from the mean.
Therefore, the low Fisher information near the origin may have little impact in practice.

\section{Discussion}
Our analysis of the monkey higher-order cortical recording shows that the SNR per neuron is very low per time bin,
such that, if the latent state changes quickly, more neurons would be required to obtain a reasonable estimate of the latent states.
These results provide a guideline for the design of experiments.
We leave the question of how to estimate the SNR for the estimation of the nonlinear dynamical state space model to future work.

The current analysis only provides an upper bound on the SNR.
Furthermore, since Fisher information necessarily depends on the assumed generative model,
the readers are encouraged to consider the assumptions when applying our results to their own experiments.

\section*{Acknowledgements}
We'd like to thank Matthew Dowling for assisting with DMFC dataset.

\vfill\pagebreak

\bibliographystyle{IEEEtran}
\bibliography{./IEEE_refs.bib,./catniplab.bib}

\begin{thebibliography}{10}
\providecommand{\url}[1]{#1}
\csname url@samestyle\endcsname
\providecommand{\newblock}{\relax}
\providecommand{\bibinfo}[2]{#2}
\providecommand{\BIBentrySTDinterwordspacing}{\spaceskip=0pt\relax}
\providecommand{\BIBentryALTinterwordstretchfactor}{4}
\providecommand{\BIBentryALTinterwordspacing}{\spaceskip=\fontdimen2\font plus
\BIBentryALTinterwordstretchfactor\fontdimen3\font minus
  \fontdimen4\font\relax}
\providecommand{\BIBforeignlanguage}[2]{{%
\expandafter\ifx\csname l@#1\endcsname\relax
\typeout{** WARNING: IEEEtran.bst: No hyphenation pattern has been}%
\typeout{** loaded for the language `#1'. Using the pattern for}%
\typeout{** the default language instead.}%
\else
\language=\csname l@#1\endcsname
\fi
#2}}
\providecommand{\BIBdecl}{\relax}
\BIBdecl

\bibitem{Degenhart2020-gu}
A.~D. Degenhart, W.~E. Bishop, E.~R. Oby, E.~C. Tyler-Kabara, S.~M. Chase,
  A.~P. Batista, and B.~M. Yu, ``\BIBforeignlanguage{en}{Stabilization of a
  brain-computer interface via the alignment of low-dimensional spaces of
  neural activity},'' \emph{\BIBforeignlanguage{en}{Nature biomedical
  engineering}}, vol.~4, no.~7, pp. 672--685, Jul. 2020.

\bibitem{Zhao2019b}
Y.~Zhao, J.~L. Yates, A.~Levi, A.~Huk, and I.~M. Park, ``Stimulus-choice
  (mis)alignment in primate area {MT},'' \emph{{PLOS} Computational Biology},
  May 2020.

\bibitem{Yu2009-qp}
B.~M. Yu, J.~P. Cunningham, G.~Santhanam, S.~I. Ryu, K.~V. Shenoy, and
  M.~Sahani, ``Gaussian-process factor analysis for low-dimensional
  single-trial analysis of neural population activity,'' \emph{Journal of
  neurophysiology}, vol. 102, no.~1, pp. 614--635, Jul. 2009.

\bibitem{Zhao2016a}
Y.~Zhao and I.~M. Park, ``Variational latent {G}aussian process for recovering
  single-trial dynamics from population spike trains,'' \emph{Neural
  Computation}, vol.~29, no.~5, May 2017.

\bibitem{Zhao2019a}
Y.~Zhao, J.~Nassar, I.~Jordan, M.~Bugallo, and I.~M. Park, ``Streaming
  variational {M}onte {C}arlo,'' \emph{IEEE Transactions on Pattern Analysis
  and Machine Intelligence}, vol.~45, no.~1, pp. 1150--1161, Feb. 2022.

\bibitem{Dowling2024b}
M.~Dowling, Y.~Zhao, and I.~M. Park, ``{eXponential} {FAmily} dynamical systems
  ({XFADS}): Large-scale nonlinear gaussian state-space modeling,'' \emph{arXiv
  [stat.ML]}, Mar. 2024.

\bibitem{Gao2015-yd}
P.~Gao and S.~Ganguli, ``\BIBforeignlanguage{en}{On simplicity and complexity
  in the brave new world of large-scale neuroscience},''
  \emph{\BIBforeignlanguage{en}{Current opinion in neurobiology}}, vol.~32, pp.
  148--155, Jun. 2015.

\bibitem{Brunel1998-pr}
N.~Brunel and J.-P. Nadal, ``Mutual information, {Fisher} information, and
  population coding,'' \emph{Neural Computation}, vol.~10, no.~7, pp.
  1731--1757, Oct. 1998.

\bibitem{Bartolo2020-zd}
R.~Bartolo, R.~C. Saunders, A.~R. Mitz, and B.~B. Averbeck,
  ``\BIBforeignlanguage{en}{{Information-Limiting} correlations in large neural
  populations},'' \emph{\BIBforeignlanguage{en}{Journal of Neuroscience}},
  vol.~40, no.~8, pp. 1668--1678, Feb. 2020.

\bibitem{Bethge2002-ji}
M.~Bethge, D.~Rotermund, and K.~Pawelzik, ``\BIBforeignlanguage{en}{Optimal
  short-term population coding: when {Fisher} information fails},''
  \emph{\BIBforeignlanguage{en}{Neural computation}}, vol.~14, no.~10, pp.
  2317--2351, Oct. 2002.

\bibitem{Beck2011-kt}
J.~Beck, V.~R. Bejjanki, and A.~Pouget, ``\BIBforeignlanguage{en}{Insights from
  a simple expression for linear {Fisher} information in a recurrently
  connected population of spiking neurons},''
  \emph{\BIBforeignlanguage{en}{Neural Computation}}, vol.~23, no.~6, pp.
  1484--1502, Jun. 2011.

\bibitem{Wei2016-mi}
X.-X. Wei and A.~A. Stocker, ``\BIBforeignlanguage{en}{Mutual information,
  {Fisher} information, and efficient coding},''
  \emph{\BIBforeignlanguage{en}{Neural computation}}, vol.~28, no.~2, pp.
  305--326, Feb. 2016.

\bibitem{Moreno-Bote2014-ie}
R.~Moreno-Bote, J.~Beck, I.~Kanitscheider, X.~Pitkow, P.~Latham, and A.~Pouget,
  ``Information-limiting correlations,'' \emph{Nature neuroscience}, vol.~17,
  no.~10, pp. 1410--1417, Oct. 2014.

\bibitem{Hurwitz2021-pf}
C.~Hurwitz, N.~Kudryashova, A.~Onken, and M.~H. Hennig,
  ``\BIBforeignlanguage{en}{Building population models for large-scale neural
  recordings: Opportunities and pitfalls},''
  \emph{\BIBforeignlanguage{en}{Current opinion in neurobiology}}, vol.~70, pp.
  64--73, Aug. 2021.

\bibitem{Pei2021b}
F.~Pei, J.~Ye, D.~Zoltowski, A.~Wu, R.~H. Chowdhury, H.~Sohn, J.~E. O'Doherty,
  K.~V. Shenoy, M.~T. Kaufman, M.~Churchland, M.~Jazayeri, L.~E. Miller,
  J.~Pillow, I.~M. Park, E.~L. Dyer, and C.~Pandarinath, ``Neural latents
  benchmark '21: Evaluating latent variable models of neural population
  activity,'' in \emph{Advances in Neural Information Processing Systems
  (NeurIPS)}, Sep. 2021, (PF and YU are co-first authors).

\bibitem{Charles2018}
A.~S. Charles, M.~Park, J.~P. Weller, G.~D. Horwitz, and J.~W. Pillow,
  ``\BIBforeignlanguage{en}{Dethroning the {F}ano factor: A flexible,
  model-based approach to partitioning neural variability},''
  \emph{\BIBforeignlanguage{en}{Neural computation}}, vol.~30, no.~4, pp.
  1012--1045, Apr. 2018.

\bibitem{Dowling2023c}
M.~Dowling, Y.~Zhao, and I.~M. Park, ``Linear time {GP}s for inferring latent
  trajectories from neural spike trains,'' in \emph{International Conference on
  Machine Learning (ICML)}, Jul. 2023.

\bibitem{Macke2011-ep}
J.~H. Macke, L.~Buesing, J.~P. Cunningham, B.~M. Yu, K.~V. Shenoy, and
  M.~Sahani, ``Empirical models of spiking in neural populations,'' in
  \emph{Advances in Neural Information Processing Systems 24}, J.~Shawe-Taylor,
  R.~S. Zemel, P.~L. Bartlett, F.~Pereira, and K.~Q. Weinberger, Eds.\hskip 1em
  plus 0.5em minus 0.4em\relax Curran Associates, Inc., 2011, pp. 1350--1358.

\bibitem{Stevenson2011-wk}
I.~H. Stevenson and K.~P. Kording, ``\BIBforeignlanguage{en}{How advances in
  neural recording affect data analysis},''
  \emph{\BIBforeignlanguage{en}{Nature neuroscience}}, vol.~14, no.~2, pp.
  139--142, Feb. 2011.

\end{thebibliography}

\end{document}